\documentclass[11pt,twocolumn,prb,superscriptaddress,reprint]{revtex4-1}
\usepackage{graphicx,amsmath,amssymb,color}
\usepackage{tabularx, ctable}
\usepackage{ulem}
\usepackage{bm}
\usepackage{siunitx}
\usepackage{gensymb}

\usepackage[breaklinks=true,colorlinks,allcolors=blue]{hyperref}

\newcommand{\be}{\begin{eqnarray}}
\newcommand{\ee}{\end{eqnarray}}

\begin{document}

\title{Electronic bistability, discontinuous switching and stochasticity \\ in a two-dimensional semiconductor} 

\author{S. Jana$^{+}$}
\affiliation{Department of Materials Science and Engineering, National University of Singapore, Singapore 117575, Singapore}
\affiliation{Institute for Functional Intelligent Materials, National University of Singapore, Singapore 117544, Singapore}

\author{M. Kravtsov$^{+}$}
\affiliation{Department of Materials Science and Engineering, National University of Singapore, Singapore 117575, Singapore}
\affiliation{Institute for Functional Intelligent Materials, National University of Singapore, Singapore 117544, Singapore}

\author{A. Ermakov}
\affiliation{School of Physics and Astronomy, University of Manchester, Manchester M13 9PL, United Kingdom}

\author{X. Zhou}
\affiliation{Institute for Functional Intelligent Materials, National University of Singapore, Singapore 117544, Singapore}

\author{A. Kudriashov}
\affiliation{Department of Materials Science and Engineering, National University of Singapore, Singapore 117575, Singapore}
\affiliation{Institute for Functional Intelligent Materials, National University of Singapore, Singapore 117544, Singapore}

\author{L. Elesin}
\affiliation{Department of Materials Science and Engineering, National University of Singapore, Singapore 117575, Singapore}
\affiliation{Institute for Functional Intelligent Materials, National University of Singapore, Singapore 117544, Singapore}

\author{X. Zhou}
\affiliation{Department of Materials Science and Engineering, National University of Singapore, Singapore 117575, Singapore}
\affiliation{Institute for Functional Intelligent Materials, National University of Singapore, Singapore 117544, Singapore}

\author{A. L. Shilov}
\affiliation{Department of Materials Science and Engineering, National University of Singapore, Singapore 117575, Singapore}
\affiliation{Institute for Functional Intelligent Materials, National University of Singapore, Singapore 117544, Singapore}

\author{D. A. Svintsov}
\affiliation{Moscow Center for Advanced Studies, Kulakova str. 20, Moscow, 123592, Russia}

\author{T. Taniguchi}
\affiliation{International Center for Materials Nanoarchitectonics, National Institute of Materials Science, Tsukuba 305-0044, Japan}

\author{K. Watanabe}
\affiliation{Research Center for Functional Materials, National Institute of Materials Science, Tsukuba 305-0044, Japan}

\author{K. S. Novoselov}
\affiliation{Department of Materials Science and Engineering, National University of Singapore, Singapore 117575, Singapore}
\affiliation{Institute for Functional Intelligent Materials, National University of Singapore, Singapore 117544, Singapore}

\author{A. Avsar}
\affiliation{Department of Materials Science and Engineering, National University of Singapore, Singapore 117575, Singapore}

\author{A. Principi}
\affiliation{School of Physics and Astronomy, University of Manchester, Manchester M13 9PL, United Kingdom}

\author{D. A. Bandurin$^{*}$}
\affiliation{Department of Materials Science and Engineering, National University of Singapore, Singapore 117575, Singapore}
\affiliation{Institute for Functional Intelligent Materials, National University of Singapore, Singapore 117544, Singapore}

\begin{abstract}

Bistability — two stable electronic states under the same bias — underlies switching and memory, but is usually absent in transistors and must be engineered through material means: doped tunnel junctions, filaments in memristors, or phase transitions. Here we demonstrate a transistor with intrinsic electronic bistability in a single chemically homogeneous crystal. In dual-gated black phosphorus, whose band gap narrows under a perpendicular electric field due to a giant Stark effect, the gates not only modulate carrier density but also reshape the band profile, forming interband tunnel junctions in the channel. Transport across the two-gate parameter space reveals competing conduction regimes — diffusive, two tunnelling channels and Zener breakdown — whose interplay produces negative differential conductance and transconductance, discontinuous switching, and hysteresis with the state set by gate history. Moreover, the switching remains intrinsically stochastic, yet statistically stable within a narrow range of gate voltages, providing an electrically programmable source of randomness. Devices based on this principle should be realisable in other two-dimensional semiconductors, opening a route to next-generation computing architectures in which nonlinearity, switching, memory and stochasticity are integrated within a single electrostatically programmable element.

\begin{center}
$^{*}$Correspondence to: dab@nus.edu.sg \\
\end{center}

\end{abstract}

\maketitle

\begin{figure*}[ht!]
  \centering\includegraphics[width=\linewidth]{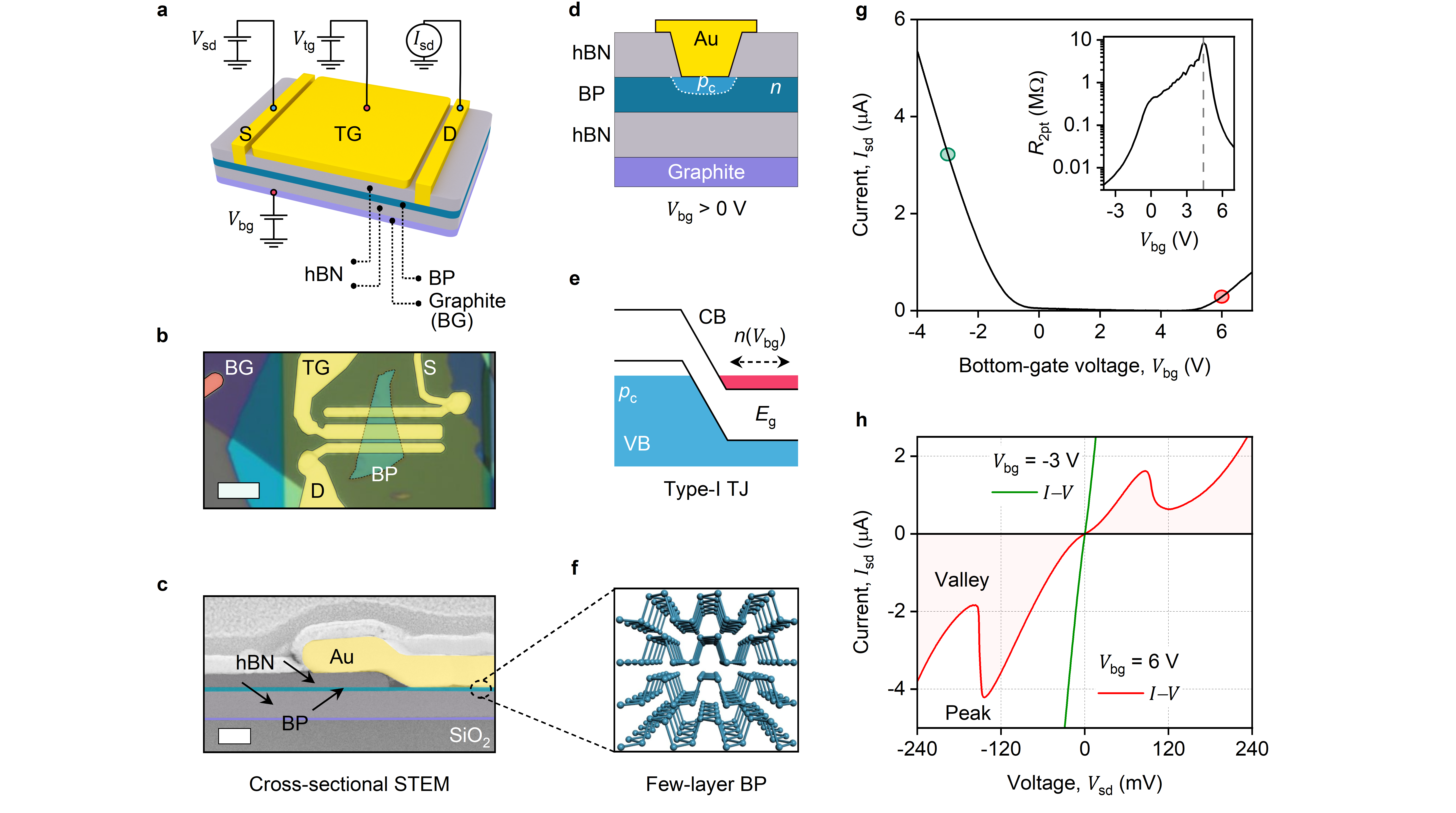}
    \caption{\textbf{Transport characteristics of the BP transistor.}
    \textbf{a,} Schematic of the BP device with the electrode configurations, \textit{i.e.}, source (S), drain (D), top gate (TG), and bottom gate (BG). \textbf{b,} A false-coloured digital photograph of the device (scale bar: $5~\mathrm{\mu m}$). \textbf{c,} False-coloured cross-sectional STEM micrograph showing the interfaces between various layers of a typical device fabricated with transferred VIA contact (scale bar: $60$~nm). Schematics showing \textbf{d,} the formation of a Type-I tunnel junction (TJ) at the Au--BP interface ($p_\mathrm{c}$--$n$) induced by the bottom gate and \textbf{e,} energy band alignment for the same. \textbf{f,} Side view of few-layer BP. \textbf{g,} Low-bias transfer characteristic of the device under the influence of $V_{\mathrm{bg}}$ ($V_\mathrm{tg} = 0$~V). The inset shows two-probe resistance ($R$ = $R_\mathrm{2pt}$) of the device. \textbf{h,} $I$--$V$ curves of the device at $V_\mathrm{bg} = -3$ and $+6$~V exhibiting superior hole injection ($p_\mathrm{c}^\mathrm{+}$--$p$ interface) and contact tunneling-mediated NDC ($p_\mathrm{c}$--$n$ interface), respectively. The values of $V_\mathrm{bg}$ for the respective $I$--$V$ characteristics are marked with circles of matching colours in \textbf{g}.
    }
	\label{Fig1}
\end{figure*}

Bistability — the coexistence of two stable electronic states under the same external bias — is a central ingredient of switches, memories, nonlinear circuits and faint light detectors.~\cite{shaw1992physics} In most solid-state devices, bistability is realised by changing the material itself: memristive, Mott and phase-change elements store information through filament formation, ionic redistribution, electronic reconstruction or structural transformation.~\cite{strukov2008missing,lanza2022memristive,song2023recent,milloch2024mott} A conceptually different route is purely electronic. In tunnel diodes and resonant-tunnelling structures, a non-monotonic current–voltage characteristic produces negative differential conductance (NDC), allowing the external circuit to stabilise distinct operating points without any permanent change in the material.~\cite{Esaki1958,chang1974resonant,shaw1992physics} This principle has enabled latches, oscillators and multivalued logic, but has remained confined to two-terminal devices whose band profiles are fixed during fabrication by doping or heteroepitaxy.~\cite{sommers1959,zaslavsky1988resonant,tanoue1988triple}

What these two-terminal devices lack is a gate. The transistor has long been the sought-after platform for realising such functionality. As the elementary unit of integrated electronics, a transistor with built-in bistability would merge the sharp switching, latch, memory effects and the amplifier into a single element: a third terminal would set, read and reconfigure the state, and circuits that pair NDC diodes with transistors — or store information in separate memory elements altogether — would collapse into one device.~\cite{lanza2022memristive,kumar2022dynamical} Transistors, however, usually do not access this regime. Their operation relies on the monotonic modulation of carrier density or barrier height by a gate: once the channel, contacts and junctions are defined, the band profile is essentially frozen, and a single homogeneous transistor offers almost no means of generating multiple transport pathways with distinct nonlinearities without relying on high-field impact ionisation or avalanche.~\cite{boudou1987hysteresis,moselund2008punch,ionescu2011tunnel,dutta2017leaky,kumar2025gallium,pazos2025synaptic} Defining a tunnel junction inside the channel does not help either. Interband tunnelling is exponentially sensitive to how abruptly the bands terminate, yet the heavy doping that forms such a junction pulls disorder-induced tails of states into the gap and leaves defects that carry current before tunnelling begins~\cite{ionescu2011tunnel,Yablonovitch}.
Electrostatic gating avoids these defects, and split gates routinely define lateral p–n homojunctions in ambipolar two-dimensional crystals, but the depletion width set by the gate dielectric rather than by a dopant profile have kept interband tunnelling out of reach. Esaki-type tunnelling in two-dimensional systems has therefore been realised only in vertical geometries, where the tunnelling distance is set by the interface itself — a direct contact or a deliberately inserted hBN barrier — and the junction by the choice of the two constituent crystals~\cite{ChhowallaNDC,RusenNDC,TaniaNDC,Mishchenko2014,Fallahazad}. Electronic bistability inside one transistor, therefore, calls for a system in which the band profile is not set at fabrication but can be reshaped in operando, so that competing conduction mechanisms can be created and tuned within the same channel.

Here we demonstrate this regime in dual-gated few-layer BP. The key enabling property is the giant Stark effect~\cite{BP-ARPES,liu2015switching,liu2017gate,deng2017efficient,tian2026reconfigurable}, through which a perpendicular electric field strongly reduces the band gap, while ambipolar transport allows the gates to define local p- and n-type regions~\cite{li2014black,chen2017widely,chen2020widely,kim2020thickness}. Together with the low disorder of few-layer BP~\cite{LiuBP,cao2015quality,li2015quantumBP,chen2015highBP,li2016quantumBP,li2019high}, these capabilities enable independently biased gates to create interband tunnel junctions within a chemically homogeneous crystal. By varying the gate voltages, we reshape the band profile to access diffusive conduction, two distinct interband-tunnelling channels, and Zener breakdown within the same device. Across these regimes, we observe NDC and negative differential transconductance (NDT), discontinuous transistor switching, and robust hysteretic electronic bistability—without dopants, heterointerfaces, conductive filaments, or phase transitions. Requiring only a strong Stark-tunable band gap and ambipolar transport, this principle should extend beyond BP to other two-dimensional semiconductors, offering a route to devices in which nonlinearity, switching, and memory-like behaviour are programmed electrostatically rather than chemically.

\textbf{Device architecture and negative differential conductance}. We fabricated our devices from an $\approx 15$~nm-thick BP flake. The device, shown in Fig.~\ref{Fig1}a,b, consists of a BP layer encapsulated by hexagonal boron nitride (hBN) and equipped with a global graphite bottom-gate and gold (Au) top-gate finger electrode. The heterostructure was assembled in an inert environment inside an argon-filled glovebox. To ensure low-resistance contacts between BP and the metal leads, gold VIAs were fabricated (Supplementary Fig.~S1) and incorporated into the top hBN slab (see Methods). This preserved the integrity of the heterostructure after removal from the glovebox for the subsequent nanofabrication steps required to connect the VIAs to the gold contact leads.~\cite{telford2018via,jung2019transferred} A typical cross-sectional scanning transmission electron microscopy (STEM) image of the device, presented in Fig.~\ref{Fig1}c, elucidates the resulting arrangement of the interfaces between the various 2D flakes and VIA contacts.

\begin{figure*}[ht!]
  \centering\includegraphics[width=1\linewidth]{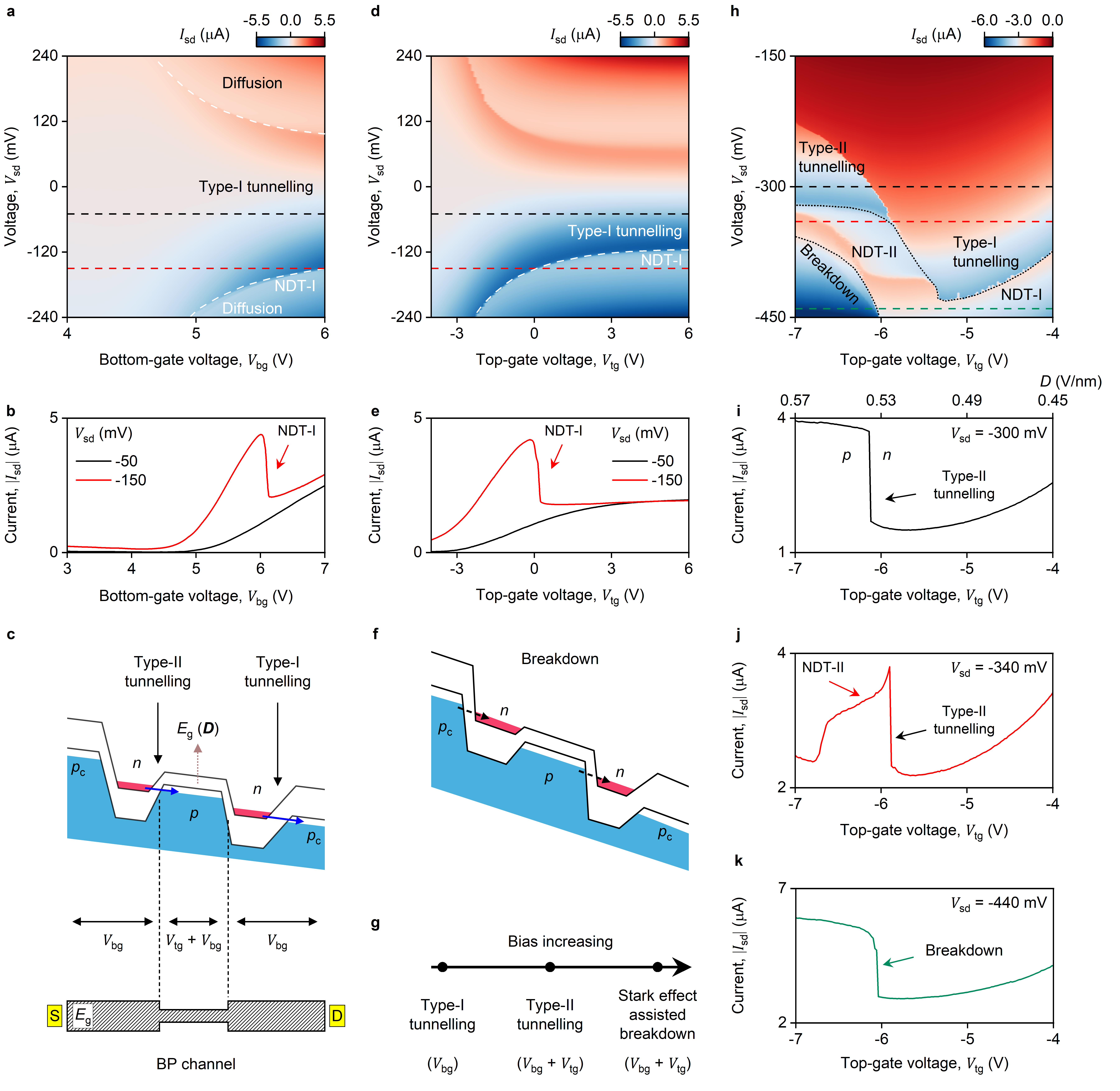}
    \caption{\textbf{Evolution of tunnel characteristics with bias and gate voltage.}
    \textbf{a,} Contour map showing the evolution of Type-I tunnelling and NDC with $V_\mathrm{bg}$ ($V_\mathrm{tg} = 0$~V). The white dashed line traces the tunnel peak current. \textbf{b,} The bottom panel shows the transistor transfer characteristics with emergence of NDT as a function of $V_\mathrm{bg}$ (along the black and red dashed lines in the map in \textbf{a}). \textbf{c,} Band alignment showing the formation of additional $p$--$n$ tunnel interfaces (Type-II TJ) along the channel mediated by the gate electrostatics. Note the Stark-effect-mediated modulation of the band gap along the channel (bottom panel). \textbf{d,} Contour map showing the evolution of tunnel-NDC with $V_\mathrm{tg}$ ($V_\mathrm{bg} = 6$~V) in low-bias regime. The highlighted dashed line shows that the values of $I_\mathrm{peak}$ are comparable for a range of $V_\mathrm{tg}$. \textbf{e,} Line traces showing emergence of NDT at the respective biases marked with dashed lines in \textbf{d}. \textbf{f,} Band alignment showing the breakdown features controlled by the gate voltage through the creation of $n$--$p$--$n$ junctions (Type-II). \textbf{g,} Distinct operational regimes achievable with the applied bias and gate voltage. \textbf{h,} Contour map showing the evolution of Type-II tunnell-NDT-breakdown with $V_\mathrm{tg}$ ($V_\mathrm{bg} = 6$~V). \textbf{i--j,} Line traces in \textbf{h} showing bias- and gate-voltage-controlled tunnelling, NDT and breakdown features.
    }
	\label{Fig2}
\end{figure*}

First, we characterised the transport properties of our BP device. Fig.~\ref{Fig1}g shows the dependence of the source--drain current, $I_\mathrm{sd}$, on the bottom-gate voltage, $V_\mathrm{bg}$, revealing a standard ambipolar transistor transfer characteristic.~\cite{li2014black} At the relatively small source--drain bias, $V_\mathrm{sd}$, used for these measurements, $I_\mathrm{sd}$ increases rapidly once the device exits the pinch-off regime, which spans $V_\mathrm{bg}=0$--$5$~V. The transfer curve is slightly shifted to positive $V_\mathrm{bg}$, indicating an unintentional $p$-doping of the channel. Importantly, the two-terminal resistance, $R=V_\mathrm{sd}/I_\mathrm{sd}$, shown in the inset of Fig.~\ref{Fig1}g, differs between the $p$- and $n$-type channel regimes. This asymmetry points to the formation of $p_\mathrm{c}$--$n$ junctions at the Au/BP interfaces, caused by charge transfer due to a large mismatch in work function mismatch, as illustrated in Fig.~\ref{Fig1}d.~\cite{buscema2014photovoltaic,pan2016monolayer} Here, $p_\mathrm{c}$ denotes contact-induced doping.

Having established the formation of contact-induced junctions at the Au/BP interfaces, we next examine how these junctions affect transport. Fig.\ref{Fig1}h presents two representative $I_\mathrm{sd}$--$V_\mathrm{sd}$ characteristics measured at $V_\mathrm{bg}=-3$ and $+6$~V. In the $p$-type regime ($V_\mathrm{bg}=-3$~V), the response is linear, with a conductance of $0.14$~mS, indicative of Ohmic contact between the Au electrodes and doped BP. This corresponds to a two-terminal resistance of $R=1/G=7.2$~k$\Omega$. In striking contrast, at $V_\mathrm{bg}=+6$~V the device exhibits a highly nonlinear $I_\mathrm{sd}$--$V_\mathrm{sd}$ characteristic. As $V_\mathrm{sd}$ is increased, $I_\mathrm{sd}$ first rises to a maximum, $I_\mathrm{peak}$, then rapidly decreases, producing a region of NDC, before reaching a minimum, $I_\mathrm{valley}$, and increasing again. The NDC appears at a relatively small bias of $V_\mathrm{sd}\approx86$~mV and has a peak-to-valley current ratio of $2.5$.

This pronounced gate tunability, from a linear Ohmic response to NDC at relatively small source--drain bias, points to Esaki-diode-like tunnelling at the BP/Au interfaces (that we later refer to as TJ-1 and TJ-2). Indeed, BP is a low-disorder, narrow-bandgap semiconductor ($\approx 300$~meV), which makes it possible to engineer spatially varying band profiles electrostatically.~\cite{li2014black,wu2018complementary,kim2020thickness,srivastava2021resonant,wu2012three,cheng2021modulation,kim2022emergence,shim2016phosphorene,abraham2020astability} In particular, the Fermi level can be driven deep into either the valence band (VB) or the conduction band (CB), enabling the formation of Esaki-type tunnel configurations (later referred to as Type-I tunnelling) through the combined action of charge transfer from the gold electrodes and the applied bottom-gate voltage, as illustrated in Figs.~\ref{Fig1}d,e. This interpretation is further supported by Fig.~\ref{Fig2}a, which maps $I_\mathrm{sd}$ as a function of $V_\mathrm{sd}$ and $V_\mathrm{bg}$. The map reveals that the NDC onset voltage is controlled by \(V_{\mathrm{bg}}\); for \(V_{\mathrm{bg}}<4.7~\mathrm{V}\), Type-I tunnelling may still occur, but the applied bias does not produce a sufficient reduction in band overlap to generate NDC within the measured \(V_{\mathrm{sd}}\) range.

\textbf{Discontinuous switching and negative differential transconductance}. A direct consequence of the gate-controlled NDC is the emergence of NDT in gate sweeps performed at fixed source--drain bias.~\cite{capasso1987negative} Fig.~\ref{Fig2}b shows $I_\mathrm{sd}$ as a function of $V_\mathrm{bg}$ for two representative values of $V_\mathrm{sd}$. At a small bias, $V_\mathrm{sd}=-50$~mV, the current increases monotonically with increasing $V_\mathrm{bg}$, as expected when the channel is driven into the electron-doped regime. By contrast, at a larger bias, $V_\mathrm{sd}=-150$~mV, $I_\mathrm{sd}$ first increases with $V_\mathrm{bg}$, reaches a pronounced maximum near $V_\mathrm{bg}\approx6$~V, and then decreases abruptly before rising again. The decreasing part of this transfer curve corresponds to negative transconductance (NDT), $g_\mathrm{m}=\partial I_\mathrm{sd}/\partial V_\mathrm{bg}<0$. This behaviour reflects the same tunnel-diode physics responsible for NDC: increasing $V_\mathrm{bg}$ changes the relative alignment of the contact-induced $p_\mathrm{c}$ region and the gate-induced $n$-type channel, driving the junction through a tunnelling condition and then out of it.

The top-gate voltage can also tune the range of $V_\mathrm{sd}$ over which NDC and NDT are observed, but through two notably different mechanisms. At positive $V_\mathrm{tg}$, the top gate has little influence on the bias at which NDC develops ($V_\mathrm{bg}= 6$~V). As $V_\mathrm{tg}$ is reduced below $0$~V, however, the NDC line progressively bends until the BP channel reaches the pinch-off state, where the Fermi level lies in the band gap. The emergence of NDT in this regime, shown in Fig.~\ref{Fig2}e, is very similar to that observed in the bottom-gate voltage scans in Fig.~\ref{Fig2}b. These observations suggest that, for $V_\mathrm{tg}>-3$~V, the top gate primarily changes the resistance of the BP channel. This changes how the applied $V_\mathrm{sd}$ is divided between the channel and the BP/Au junctions, thereby shifting the voltage drop across the tunnelling interface and modifying the condition for Type-I tunnelling (see Supplementary Fig.~S2 and S3 for details).

When $V_\mathrm{tg}$ is made sufficiently negative, the top gate qualitatively changes the band profile of the device. The bottom-gated parts of the BP channel remain electron-doped, whereas the region underneath the top gate is depleted and eventually becomes hole-doped. The device therefore acquires an electrostatically defined $n$--$p$--$n$ profile in the BP channel, in addition to the contact-induced $p_\mathrm{c}$--$n$ junctions at the Au/BP interfaces. Importantly, the top and bottom gates also generate a vertical displacement field, $\textit{D}$, in the top-gated region (for $V_\mathrm{tg} = -6$~V, $\textit{D}\approx 0.5$~V/nm, considering $\epsilon_\mathrm{r} = 3.9$ for hBN). This field reduces the local BP band gap,~\cite{liu2015switching,liu2017gate,deng2017efficient,chen2017widely,chen2020widely,tian2026reconfigurable} $E_\mathrm{g}$(\textit{D}), as illustrated in Fig.~\ref{Fig2}c, and thereby lowers the energy cost for aligning occupied VB states in the $p$-type region with empty conduction-band states in the adjacent $n$-type regions. This creates a second class of Esaki-type junctions inside the BP channel itself, labelled as TJ-3 and TJ-4. We refer to tunnelling across these electrostatically defined $n$--$p$ interfaces as Type-II tunnelling, to distinguish it from Type-I tunnelling at the BP/Au contact. Esaki-type NDC at a gate-defined junction in a chemically homogeneous two-dimensional crystal has not been reported previously; the closest prior observation in ambipolar BP resolved only a precursor to NDC~\cite{kim2022emergence}.

\begin{figure*}[ht!]
  \centering\includegraphics[width=1\linewidth]{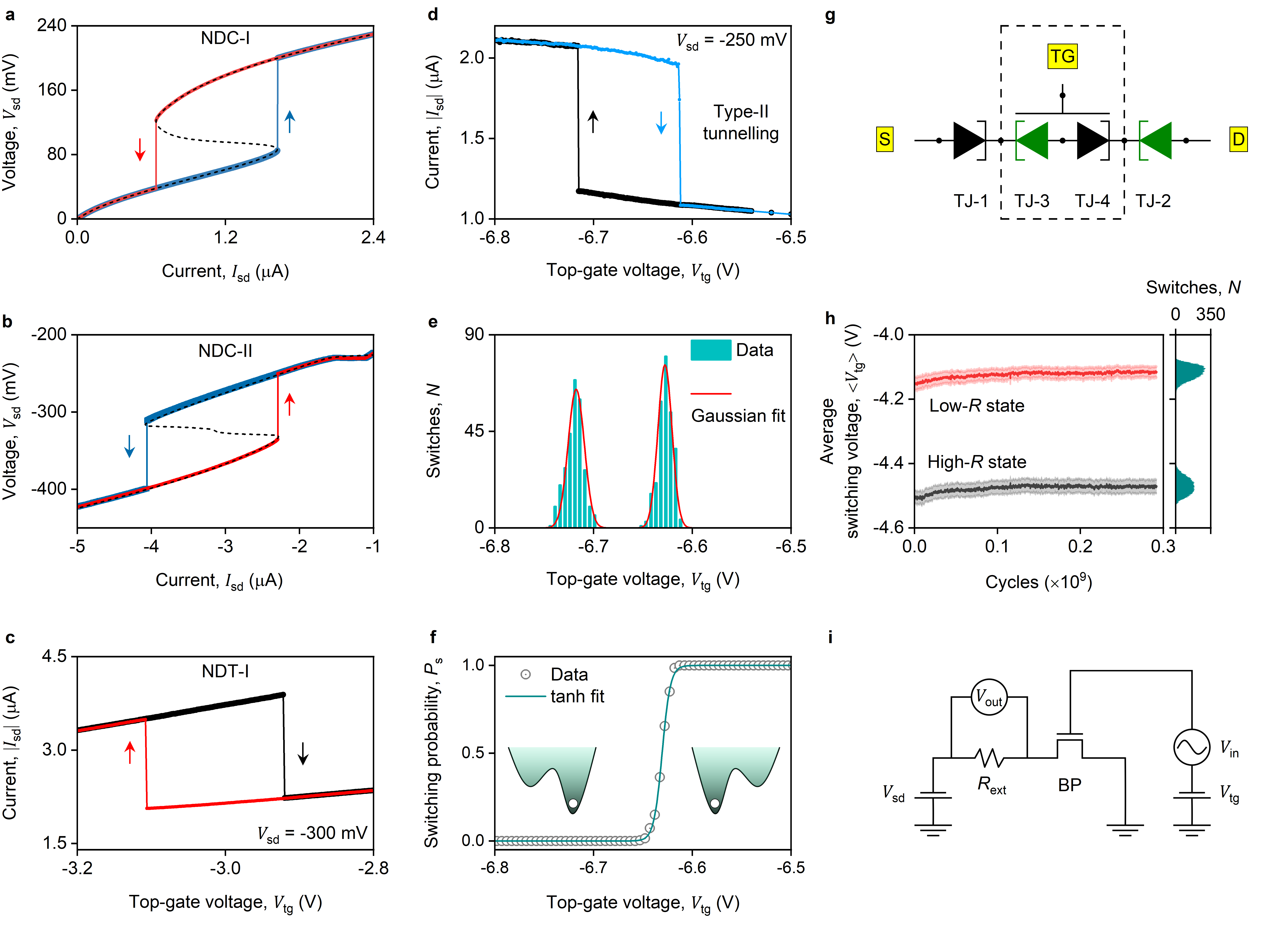}
    \caption{\textbf{Manifold electronic bistability in BP.}
    \textbf{a,} A typical voltage hysteresis in the current bias regime for Type-I NDC ($V_\mathrm{tg}= 0$~V). Blue and red traces are increasing and decreasing sweeps; the dashed line is the unstable branch inferred from the voltage-biased characteristics. \textbf{b,} The same measurement in the gate-defined Type-II NDC ($V_\mathrm{tg}= -7$~V). \textbf{c,} Transfer characteristics exhibiting hysteresis in \textbf{c,} Type-I NDT and \textbf{d,} Type-II tunnelling. Arrows indicate the direction of the scan and the jumps in voltage and current, respectively. \textbf{e,} Distribution of the gate voltage at which the switch in \textbf{d} occurs, compiled over 300 cycles, with Gaussian fits ($\sigma = 8.2$~mV (forward, LHS) and $7.2$~mV (backwards, RHS). \textbf{f,} Switching probability extracted from \textbf{e} (backward sweeps) with a tanh fit; insets sketch the double-well effective potential at the marked gate voltages. \textbf{g,} Equivalent circuit: contact-defined junctions TJ-1 and TJ-2 in series with the gate-defined junctions TJ-3 and TJ-4; junctions forward-biased for ($V_\mathrm{sd}< 0$~V) are green. \textbf{h,} Mean switching voltage of the low- and high-resistance branches over $\sim3\times10^8$ cycles. The shaded region represents FWHMs of the switching gate voltage. The side panel shows the switches for the last sampling. \textit{i,} Circuit used for \textbf{h}: the current is read as the voltage $V_\mathrm{out}$ across $R_\mathrm{ext}$ while a sinusoidal $V_\mathrm{in}$ is superimposed on the DC top-gate voltage (see Methods).
    }
	\label{Fig3}
\end{figure*}

This picture explains the richer transport map observed for $V_\mathrm{tg}<-5$~V in Fig.~\ref{Fig2}h. At small $|V_\mathrm{sd}|$, the top-gated region is close to pinch-off, and the current is strongly suppressed because the Fermi level lies in the gap and the interband tunnelling window is closed. As $|V_\mathrm{sd}|$ is increased, the bands tilt along the channel, and the VB in the top-gated $p$ region becomes aligned with the CB in the neighbouring $n$ region. This opens the Type-II tunnelling channel and produces an abrupt, discontinuous increase in current, as seen in the fixed-bias trace at $V_\mathrm{sd}=-300$~mV in Fig.~\ref{Fig2}i. At larger bias, the tunnelling window itself becomes strongly gate dependent: sweeping $V_\mathrm{tg}$ first improves and then suppresses the band overlap, giving rise to an additional negative-transconductance feature, NDT-II, shown in Fig.~\ref{Fig2}j. 

The tunnelling phase map distinguishes the two tunnel channels and separates the operational regimes in bias and gate voltage. Beyond NDT-I and NDT-II, at still higher bias, the current jumps once more (Fig.~\ref{Fig2}k), and this jump bounds a further region of the map. The forward-biased junctions cannot produce it: at this bias, they have already passed the valley, and the post-tunnel diffusion model accounts for neither the abruptness nor the magnitude of the step. We attribute the jump instead to the reverse-biased junctions, TJ-1 and TJ-4, at which the interband window does not close with increasing bias but widens, so that these junctions become progressively more transparent to Zener tunnelling~\cite{zener1934theory}. Two factors open this window: the bands tilt further as the reverse bias grows, and the Stark-narrowed gap of the top-gated region lowers the barrier to be traversed. Once the resulting breakdown current exceeds the diffusion current of the forward-biased junctions, it sets the total current, and the device enters a standalone breakdown phase whose threshold is set electrostatically by the gates rather than by the band gap of the unbiased crystal --- reached here at a longitudinal field of only $<1$~kV/cm.

\textbf{Transistor latching, bistability and stochastic switching}. The NDC regimes described above provide a direct route to electronic bistability. Under current-biased operation, the NDC characteristic permits two stable voltage states at the same imposed current, separated by an unstable branch. This behaviour is demonstrated for the Type-I and Type-II regimes in Figs.~\ref{Fig3}a and b, respectively. During a forward current sweep, the device follows one stable branch until it reaches a critical current, where the voltage switches discontinuously to the other branch. When the sweep direction is reversed, the device remains latched in this state until a second threshold is reached, producing a pronounced hysteresis loop. The opposite switching polarity and different voltage scales of NDC-I and NDC-II reflect their distinct band-alignment configurations, but both exhibit the same defining features of bistability: branch termination, discontinuous switching and history-dependent state selection.

The same instability appears in the voltage-biased transistor transfer characteristics as NDT and gate-controlled latching. Figs.~\ref{Fig3}c and \ref{Fig3}d show $I_\mathrm{sd}(V_\mathrm{tg})$ measured at fixed $V_\mathrm{sd}$ across the Type-I and Type-II tunnelling regimes. Sweeping $V_\mathrm{tg}$ continuously reshapes the band-alignment profile and changes how the applied source--drain voltage is distributed across the junctions and the intervening BP channel. In the Type-I configuration, the voltage is shared predominantly between the two contact-associated tunnel junctions, TJ-1 and TJ-2, and the channel, such that
$V_\mathrm{sd} \approx V_{\mathrm{TJ\mbox{-}1}} + V_\mathrm{ch} + V_{\mathrm{TJ\mbox{-}2}}$. In the Type-II configuration, two additional gate-defined junctions, TJ-3 and TJ-4, form inside the BP channel and are forward-bias coupled, giving
$V_\mathrm{sd} \approx V_{\mathrm{TJ\mbox{-}1}} + V_{\mathrm{TJ\mbox{-}3}} + V_{\mathrm{TJ\mbox{-}4}} + V_{\mathrm{TJ\mbox{-}2}},$
as illustrated in Fig.~\ref{Fig3}g. Because all junctions are connected in series, the current through one junction determines the voltage available to the others. Within the NDC range, this self-consistent redistribution of voltage supports two stable current states at the same $V_\mathrm{tg}$ and $V_\mathrm{sd}$. The device therefore remains ON in one branch until that solution loses stability, whereupon the internal voltage distribution reorganises abruptly and $I_\mathrm{sd}$ jumps to the other branch. Upon reversing the gate sweep, the device stays latched to this new state until the second stability boundary is crossed. Consequently, within the hysteretic interval, the same gate voltage can correspond to either a high-current or a low-current state depending on the preceding sweep history. The transfer characteristic thus acquires a memory-like response even though no material state is written or retained: the apparent memory originates from a gate-programmed electronic instability of the tunnel junction. 

To quantify the switching further, we repeated the gate sweep for 300 cycles and extracted the switching voltage for each transition. The resulting histograms, shown in Fig.~\ref{Fig3}e, reveal stochastic cycle-to-cycle variations around well-defined mean thresholds resembling how thermal fluctuations in superconducting nanowires affect the switching current. The microscopic origin of these fluctuations remain to be understood, but their temperature-independent width and insensitivity to the gate-voltage source disfavour simple thermal activation and source-induced noise; plausible contributions include nonequilibrium tunnelling-current fluctuations which can be strongly amplified near the termination of a bistable branch. Importantly, although individual switching events are probabilistic, the two current states, the hysteresis window and the overall switching distributions remain stable over repeated cycling. To assess this stability over extended operation, we applied a combined DC+AC signal to the transistor gate, as illustrated in Fig.~\ref{Fig3}i, and monitored the switching over almost $3\times10^8$ cycles. No detectable degradation was observed, while both the mean switching voltages and their statistical spread remained nearly unchanged (Fig.~\ref{Fig3}h).

The stability of the switching statistics enables the device to operate as an electrically programmable probabilistic bit~\cite{Borders2019,Park2026}. Integrating the switching-voltage histograms in Fig.~\ref{Fig3}e yields the cumulative switching probability, $P_{\mathrm{s}}$, shown in Fig.~\ref{Fig3}f. Rather than changing abruptly, $P_{\mathrm{s}}$ evolves continuously from zero to unity across a narrow gate-voltage interval and is well described by a hyperbolic-tangent function, resembling the activation characteristic of a binary stochastic neuron. In the circuit shown in Fig.~\ref{Fig3}i, the DC gate voltage sets the operating point, while $V_{\mathrm{in}}$ controls the amplitude of the AC gate excitation. At small amplitudes, the switching threshold is not reached and $P_{\mathrm{s}}\approx 0$; at sufficiently large amplitudes, switching occurs during every cycle and $P_{\mathrm{s}}\approx 1$. Within the intermediate stochastic window, the switching probability varies continuously with the drive amplitude, such that the output voltage, $V_{\mathrm{out}}$, switches between the two stable states with an electrically controlled probability. The BP transistor therefore combines controlled randomness with reproducible operating characteristics, making it attractive for probabilistic bits and sampling-based hardware~\cite{Borders2019,Park2026}. Unlike filamentary memristive devices~\cite{lanza2022memristive}, in which stochasticity can be accompanied by structural rearrangement, threshold drift, and cycle-to-cycle degradation, the randomness here arises from a reversible electrostatic instability in a chemically homogeneous crystal.

\textbf{Conclusions and outlook}.
We have demonstrated a transistor in which bistability and discontinuous switching arise from a reversible electronic instability rather than a material transformation. In dual-gated few-layer BP, gate-defined and contact-associated tunnel junctions are coupled in series, producing negative differential conductance and transconductance, latching and two stable states under the same bias. Individual switching thresholds fluctuate within reproducible distributions over repeated cycles. The device therefore combines nonlinearity, memory-like behaviour and controlled randomness within one chemically homogeneous semiconductor.

Gate-tunable NDC, bistability and stochastic switching within one transistor open up several device opportunities. Its two stable branches can form a compact latch or memory element, with the gate providing a third terminal for writing, reading and reconfiguring the state without material transformation. Capacitive loading could convert the NDC regime into tunable relaxation oscillators and stochastic spiking elements for neuromorphic circuits. The reproducible switching distributions further suggest probabilistic bits and hardware random number generators controlled electrostatically rather than through magnetic, ionic or structural switching. Networks of such devices could combine sampling, nonlinear activation and short-term memory for probabilistic accelerators, reconfigurable logic and adaptive mixed-signal circuits.

\textbf{Materials and Methods}

\textbf{Device fabrication}. Our devices were fabricated using a dry-transfer technique described elsewhere.~\cite{purdie2018cleaning} The device consisted of a thin BP flake encapsulated by hBN and electrically accessed through transferred gold VIA contacts.~\cite{telford2018via,jung2019transferred,arora2023fully} Bulk BP crystals were purchased from HQ Graphene, exfoliated, and stacked together with hBN and graphite inside an Ar-filled glovebox with $\mathrm{O}_2$ and $\mathrm{H_2O}$ levels below~1~ppm. A PC/PDMS stamp was used to pick up an hBN flake containing the VIA contacts, followed by a thin BP flake at $65~^\circ\mathrm{C}$. The completed stack was released onto a highly insulating SiO$_2$/i-Si substrate ($285/5250$~nm, $>20$~k$\Omega\cdot$cm), after which the residual polymer was removed in dichloromethane and rinsed in isopropyl alcohol. A final lithography step defined the contact pads, and Au/Ti (85/5~nm) was deposited. The channel length is $\sim3.8~\mu$m. Devices were kept inside the glovebox between fabrication steps and prior to measurement.

\textbf{Transport measurements}. The transport measurements were performed in a two-probe configuration in a variable-temperature cryostat. Instrumentation included SR860/830 lock-in amplifiers, a Lakeshore M81-SSM and a Keithley SM2614B sourcemeter.

\textbf{Switching statistics}. Hysteretic switching was characterised using a fast current-readout scheme. A resistor $R_\mathrm{ext} = 2$~$\mathrm{k\Omega}$, much smaller than the device resistance, was placed in series with the channel, and the voltage drop $V_\mathrm{out}$ across it was digitised with a National Instruments cDAQ-9174, providing a direct measure of the circuit current. The top gate was driven by a sinusoidal waveform $V_\mathrm{in}$ superimposed on a DC offset $V_\mathrm{tg}$, while $I(t)$ was sampled continuously at $\sim$102~kSa/s. Switching events were identified by numerically differentiating the $I(t)$ traces and registering excursions of $\mathrm{d}I/\mathrm{d}t$ above a fixed threshold, from which the switching statistics were compiled.

\vspace{1em}

\textbf{Acknowledgements}.
This work was supported by A*STAR under its Young Individual Research Grant (award M22K3c0106).

\bibliography{Bibliography.bib}

@article{BP-ARPES,
author = {Jimin Kim  and Seung Su Baik  and Sae Hee Ryu  and Yeongsup Sohn  and Soohyung Park  and Byeong-Gyu Park  and Jonathan Denlinger  and Yeonjin Yi  and Hyoung Joon Choi  and Keun Su Kim },
title = {Observation of tunable band gap and anisotropic Dirac semimetal state in black phosphorus},
journal = {Science},
volume = {349},
number = {6249},
pages = {723-726},
year = {2015},
publisher = {American Association for the Advancement of Science},
doi = {10.1126/science.aaa6486}
}

@article{LiuBP,
author = {Liu, Han and Neal, Adam T. and Zhu, Zhen and Luo, Zhe and Xu, Xianfan and Tománek, David and Ye, Peide D.},
title = {Phosphorene: An Unexplored 2D Semiconductor with a High Hole Mobility},
journal = {ACS Nano},
volume = {8},
number = {4},
pages = {4033-4041},
year = {2014},
publisher = {ACS},
doi ={10.1021/nn501226z}
}

@article{jung2019transferred,
  title={Transferred via contacts as a platform for ideal two-dimensional transistors},
  author={Jung, Younghun and Choi, Min Sup and Nipane, Ankur and Borah, Abhinandan and Kim, Bumho and Zangiabadi, Amirali and Taniguchi, Takashi and Watanabe, Kenji and Yoo, Won Jong and Hone, James and others},
  journal={Nature Electronics},
  volume={2},
  number={5},
  pages={187--194},
  year={2019},
  publisher={Nature Publishing Group UK London},
  doi={10.1038/s41928-019-0245-y}
}

@article{telford2018via,
  title={Via method for lithography free contact and preservation of 2D materials},
  author={Telford, Evan J and Benyamini, Avishai and Rhodes, Daniel and Wang, Da and Jung, Younghun and Zangiabadi, Amirali and Watanabe, Kenji and Taniguchi, Takashi and Jia, Shuang and Barmak, Katayun and others},
  journal={Nano letters},
  volume={18},
  number={2},
  pages={1416--1420},
  year={2018},
  publisher={ACS Publications},
  doi={10.1021/acs.nanolett.7b05161}
}

@article{li2019high,
  title={High-speed black phosphorus field-effect transistors approaching ballistic limit},
  author={Li, Xuefei and Yu, Zhuoqing and Xiong, Xiong and Li, Tiaoyang and Gao, Tingting and Wang, Runsheng and Huang, Ru and Wu, Yanqing},
  journal={Science advances},
  volume={5},
  number={6},
  pages={eaau3194},
  year={2019},
  publisher={American Association for the Advancement of Science},
  doi={10.1126/sciadv.aau3194}
}

@article{buscema2014photovoltaic,
  title={Photovoltaic effect in few-layer black phosphorus PN junctions defined by local electrostatic gating},
  author={Buscema, Michele and Groenendijk, Dirk J and Steele, Gary A and Van Der Zant, Herre Sj and Castellanos-Gomez, Andres},
  journal={Nature communications},
  volume={5},
  number={1},
  pages={4651},
  year={2014},
  publisher={Nature Publishing Group UK London},
  doi={10.1038/ncomms5651}
}

@article{Fallahazad,
    author = {Fallahazad, Babak and Lee, Kayoung and Kang, Sangwoo and Xue, Jiamin and Larentis, Stefano and Corbet, Christopher and Kim, Kyounghwan and Movva, Hema C. P. and Taniguchi, Takashi and Watanabe, Kenji and Register, Leonard F. and Banerjee, Sanjay K. and Tutuc, Emanuel},
    title = {Gate-Tunable Resonant Tunneling in Double Bilayer
Graphene Heterostructures},
    journal = {Nano Letters},
    volume = {15},
    number = {1},
    pages = {428-433},
    year = {2014},
}

@Article{Mishchenko2014,
author={Mishchenko, A.
and Tu, J. S.
and Cao, Y.
and Gorbachev, R. V.
and Wallbank, J. R.
and Greenaway, M. T.
and Morozov, V. E.
and Morozov, S. V.
and Zhu, M. J.
and Wong, S. L.
and Withers, F.
and Woods, C. R.
and Kim, Y.-J.
and Watanabe, K.
and Taniguchi, T.
and Vdovin, E. E.
and Makarovsky, O.
and Fromhold, T. M.
and Fal'ko, V. I.
and Geim, A. K.
and Eaves, L.
and Novoselov, K. S.},
title={Twist-controlled resonant tunnelling in graphene/boron nitride/graphene heterostructures},
journal={Nature Nanotechnology},
year={2014},
volume={9},
number={10},
pages={808-813},
}

@article{srivastava2021resonant,
  title={Resonant tunnelling diodes based on twisted black phosphorus homostructures},
  author={Srivastava, Pawan Kumar and Hassan, Yasir and de Sousa, Duarte JP and Gebredingle, Yisehak and Joe, Minwoong and Ali, Fida and Zheng, Yang and Yoo, Won Jong and Ghosh, Subhasis and Teherani, James T and others},
  journal={Nature Electronics},
  volume={4},
  number={4},
  pages={269--276},
  year={2021},
  publisher={Nature Publishing Group UK London},
  doi={10.1038/s41928-021-00549-1}
}

@Article{Borders2019,
author={Borders, William A. and Pervaiz, Ahmed Z. and Fukami, Shunsuke and Camsari, Kerem Y. and Ohno, Hideo and Datta, Supriyo},
title={Integer factorization using stochastic magnetic tunnel junctions},
journal={Nature},
year={2019},
volume={573},
number={7774},
pages={390-393},
publisher={Nature Publishing Group UK London},
  doi={10.1038/s41586-019-1557-9}
}

@Article{Park2026,
author={Park, Jun-Young and Lee, Jae-Hyun and Lee, Jeong-Min and Park, Ki-Nam and Kim, Sungho and Han, Joon-Kyu},
title={CMOS compatible probabilistic computing hardware with cointegrated reconfigurable p-bits and synapse arrays},
journal={Nature Communications},
year={2026},
volume={17},
number={1},
pages={5110},
publisher={Nature Publishing Group UK London},
  doi={10.1038/s41467-026-71906-x}
}

@article{kim2020thickness,
  title={Thickness-controlled black phosphorus tunnel field-effect transistor for low-power switches},
  author={Kim, Seungho and Myeong, Gyuho and Shin, Wongil and Lim, Hongsik and Kim, Boram and Jin, Taehyeok and Chang, Sungjin and Watanabe, Kenji and Taniguchi, Takashi and Cho, Sungjae},
  journal={Nature nanotechnology},
  volume={15},
  number={3},
  pages={203--206},
  year={2020},
  publisher={Nature Publishing Group UK London},
  doi={10.1038/s41565-019-0623-7}
}

@article{kim2022emergence,
  title={Emergence of quantum tunneling in ambipolar black phosphorus multilayers without heterojunctions},
  author={Kim, Yeeun and Kim, Chulmin and Kim, Soo Yeon and Lee, Byung Chul and Seo, Youkyung and Cho, Hyeran and Kim, Gyu-Tae and Joo, Min-Kyu},
  journal={Advanced Functional Materials},
  volume={32},
  number={13},
  pages={2110391},
  year={2022},
  publisher={Wiley Online Library},
  doi={10.1002/adfm.202110391}
}

@article{pan2016monolayer,
  title={Monolayer phosphorene--metal contacts},
  author={Pan, Yuanyuan and Wang, Yangyang and Ye, Meng and Quhe, Ruge and Zhong, Hongxia and Song, Zhigang and Peng, Xiyou and Yu, Dapeng and Yang, Jinbo and Shi, Junjie and others},
  journal={Chemistry of Materials},
  volume={28},
  number={7},
  pages={2100--2109},
  year={2016},
  publisher={ACS Publications},
  doi={10.1021/acs.chemmater.5b04899}
}

@article{li2014black,
  title={Black phosphorus field-effect transistors},
  author={Li, Likai and Yu, Yijun and Ye, Guo Jun and Ge, Qingqin and Ou, Xuedong and Wu, Hua and Feng, Donglai and Chen, Xian Hui and Zhang, Yuanbo},
  journal={Nature nanotechnology},
  volume={9},
  number={5},
  pages={372--377},
  year={2014},
  publisher={Nature Publishing Group UK London},
  doi={10.1038/nnano.2014.35}
}

@article{deng2017efficient,
  title={Efficient electrical control of thin-film black phosphorus bandgap},
  author={Deng, Bingchen and Tran, Vy and Xie, Yujun and Jiang, Hao and Li, Cheng and Guo, Qiushi and Wang, Xiaomu and Tian, He and Koester, Steven J and Wang, Han and others},
  journal={Nature communications},
  volume={8},
  number={1},
  pages={14474},
  year={2017},
  publisher={Nature Publishing Group UK London},
  doi={10.1038/ncomms14474}
}

@article{chen2017widely,
  title={Widely tunable black phosphorus mid-infrared photodetector},
  author={Chen, Xiaolong and Lu, Xiaobo and Deng, Bingchen and Sinai, Ofer and Shao, Yuchuan and Li, Cheng and Yuan, Shaofan and Tran, Vy and Watanabe, Kenji and Taniguchi, Takashi and others},
  journal={Nature communications},
  volume={8},
  number={1},
  pages={1672},
  year={2017},
  publisher={Nature Publishing Group UK London},
  doi={10.1038/s41467-017-01978-3}
}

@article{wu2018complementary,
  title={Complementary black phosphorus tunneling field-effect transistors},
  author={Wu, Peng and Ameen, Tarek and Zhang, Huairuo and Bendersky, Leonid A and Ilatikhameneh, Hesameddin and Klimeck, Gerhard and Rahman, Rajib and Davydov, Albert V and Appenzeller, Joerg},
  journal={ACS nano},
  volume={13},
  number={1},
  pages={377--385},
  year={2018},
  publisher={ACS Publications},
  doi={10.1021/acsnano.8b06441}
}

@article{Esaki1958,
  author={Esaki, Leo},
  title={{New Phenomenon in Narrow Germanium p-n Junctions}},
  journal={{Physical Review}},
  volume={109},
  issue={2},
  pages={603--604},
  year={1958},
  publisher={American Physical Society},
  doi={10.1103/PhysRev.109.603}
}

@article{ionescu2011tunnel,
  title={Tunnel field-effect transistors as energy-efficient electronic switches},
  author={Ionescu, Adrian M and Riel, Heike},
  journal={Nature},
  volume={479},
  number={7373},
  pages={329--337},
  year={2011},
  publisher={Nature Publishing Group UK London},
  doi={10.1038/nature10679}
}

@ARTICLE{Yablonovitch,
  author={Agarwal, Sapan and Yablonovitch, Eli},
  journal={IEEE Transactions on Electron Devices}, 
  title={Band-Edge Steepness Obtained From Esaki/Backward Diode Current–Voltage Characteristics}, 
  year={2014},
  volume={61},
  number={5},
  pages={1488-1493},
  publisher={IEEE},
  doi={10.1109/TED.2014.2312731}
  }

@article{cheng2021modulation,
  title={Modulation of negative differential resistance in black phosphorus transistors},
  author={Cheng, Ruiqing and Yin, Lei and Hu, Rui and Liu, Huijun and Wen, Yao and Liu, Chuansheng and He, Jun},
  journal={Advanced Materials},
  volume={33},
  number={25},
  pages={2008329},
  year={2021},
  publisher={Wiley Online Library},
  doi={10.1002/adma.202008329}
}

@article{chang1974resonant,
  title     = {Resonant tunneling in semiconductor double barriers},
  author    = {Chang, L. L. and Esaki, L. and Tsu, R.},
  journal   = {Applied Physics Letters},
  volume    = {24},
  number    = {12},
  pages     = {593--595},
  year      = {1974},
  doi       = {10.1063/1.1655067}
}

@article{purdie2018cleaning,
  title={Cleaning interfaces in layered materials heterostructures},
  author={Purdie, D Gꎬ and Pugno, NM and Taniguchi, T and Watanabe, K and Ferrari, AC and Lombardo, Antonio},
  journal={Nature communications},
  volume={9},
  number={1},
  pages={5387},
  year={2018},
  publisher={Nature Publishing Group UK London},
  doi={10.1038/s41467-018-07558-3}
}

@article{arora2023fully,
  title={Fully Encapsulated and Stable Black Phosphorus Field-Effect Transistors},
  author={Arora, Himani and Fekri, Zahra and Vekariya, Yagnika Nandlal and Chava, Phanish and Watanabe, Kenji and Taniguchi, Takashi and Helm, Manfred and Erbe, Artur},
  journal={Advanced Materials Technologies},
  volume={8},
  number={2},
  pages={2200546},
  year={2023},
  publisher={Wiley Online Library},
  doi = {10.1002/admt.202200546}
}

@article{chen2020widely,
  title={Widely tunable mid-infrared light emission in thin-film black phosphorus},
  author={Chen, Chen and Lu, Xiaobo and Deng, Bingchen and Chen, Xiaolong and Guo, Qiushi and Li, Cheng and Ma, Chao and Yuan, Shaofan and Sung, Eric and Watanabe, Kenji and others},
  journal={Science advances},
  volume={6},
  number={7},
  pages={eaay6134},
  year={2020},
  publisher={American Association for the Advancement of Science},
  doi = {10.1126/sciadv.aay6134}
}

@article{liu2015switching,
  title={Switching a normal insulator into a topological insulator via electric field with application to phosphorene},
  author={Liu, Qihang and Zhang, Xiuwen and Abdalla, LB and Fazzio, Adalberto and Zunger, Alex},
  journal={Nano letters},
  volume={15},
  number={2},
  pages={1222--1228},
  year={2015},
  publisher={ACS Publications},
  doi = {10.1021/nl5043769}
}

@article{capasso1987negative,
  title={Negative transconductance resonant tunneling field-effect transistor},
  author={Capasso, Federico and Sen, Susanta and Cho, Alfred Y},
  journal={Applied physics letters},
  volume={51},
  number={7},
  pages={526--528},
  year={1987},
  publisher={American Institute of Physics},
  doi = {10.1063/1.98387},
}

@article{zener1934theory,
  title={A theory of the electrical breakdown of solid dielectrics},
  author={Zener, Clarence},
  journal={Proceedings of the Royal Society of London. Series A, Containing Papers of a Mathematical and Physical Character},
  volume={145},
  number={855},
  pages={523--529},
  year={1934},
  publisher={The Royal Society London},
  doi = {10.1098/rspa.1934.0116}
}

@article{sommers1959,
  author={H. S. Sommers, Jr.},
  title={Tunnel Diodes as High‐Frequency Devices},
  journal={Proceedings of the IRE},
  volume={47},
  number={7},
  pages={1201--1206},
  year={1959},
  doi={10.1109/JRPROC.1959.287351}
}

@article{zaslavsky1988resonant,
  title={Resonant tunneling and intrinsic bistability in asymmetric double-barrier heterostructures},
  author={Zaslavsky, A and Goldman, VJ and Tsui, DC and Cunningham, JE},
  journal={Applied physics letters},
  volume={53},
  number={15},
  pages={1408--1410},
  year={1988},
  publisher={American Institute of Physics},
  doi={10.1063/1.99956}
}

@article{ChhowallaNDC,
    author = {Kim, Jung Ho and Sarkar, Soumya and Wang, Yan and Taniguchi, Takashi and Watanabe, Kenji and Chhowalla, Manish},
    title = {Room Temperature
Negative Differential Resistance
with High Peak Current in MoS2/WSe2 Heterostructures},
    journal = {Nano Letters},
    volume = {24},
    number = {8},
    pages = {2561-2566},
    year = {2024},
    month = {02},
}

@article{kumar2025gallium,
  title={Gallium nitride multichannel devices with latch-induced sub-60-mV-per-decade subthreshold slopes for radiofrequency applications},
  author={Kumar, Akhil S and Dalcanale, Stefano and Uren, Michael J and Pomeroy, James W and Smith, Matthew D and Parke, Justin A and Howell, Robert S and Kuball, Martin},
  journal={Nature Electronics},
  pages={1--8},
  year={2025},
  publisher={Nature Publishing Group UK London},
  doi={10.1038/s41928-025-01391-5}
}

@article{TaniaNDC,
    author = {Roy, Tania and Tosun, Mahmut and Cao, Xi and Fang, Hui and Lien, Der-Hsien and Zhao, Peida and Chen, Yu-Ze and Chueh, Yu-Lun and Guo, Jing and Javey, Ali},
    title = {Dual-Gated MoS2/WSe2 van der Waals Tunnel Diodes and Transistors},
    journal = {ACS Nano},
    volume = {9},
    number = {2},
    pages = {2071-2079},
    year = {2015},
}

@article{RusenNDC,
    author = {Yan, Rusen and Fathipour, Sara and Han, Yimo and Song, Bo and Xiao, Shudong and Li, Mingda and Ma, Nan and Protasenko, Vladimir and Muller, David A. and Jena, Debdeep and Xing, Huili Grace},
    title = {Esaki Diodes in van der Waals Heterojunctions with
Broken-Gap Energy Band Alignment},
    journal = {Nano Letters},
    volume = {15},
    number = {9},
    pages = {5791-5798},
    year = {2015},
    month = {07},
}

@article{li2016quantumBP,
  title={Quantum Hall effect in black phosphorus two-dimensional electron system},
  author={Li, Likai and Yang, Fangyuan and Ye, Guo Jun and Zhang, Zuocheng and Zhu, Zengwei and Lou, Wenkai and Zhou, Xiaoying and Li, Liang and Watanabe, Kenji and Taniguchi, Takashi and others},
  journal={Nature nanotechnology},
  volume={11},
  number={7},
  pages={593--597},
  year={2016},
  publisher={Nature Publishing Group UK London},
  doi={10.1038/nnano.2016.42}
}

@article{li2015quantumBP,
  title={Quantum oscillations in a two-dimensional electron gas in black phosphorus thin films},
  author={Li, Likai and Ye, Guo Jun and Tran, Vy and Fei, Ruixiang and Chen, Guorui and Wang, Huichao and Wang, Jian and Watanabe, Kenji and Taniguchi, Takashi and Yang, Li and others},
  journal={Nature nanotechnology},
  volume={10},
  number={7},
  pages={608--613},
  year={2015},
  publisher={Nature Publishing Group UK London},
  doi={10.1038/nnano.2015.91}
}

@article{chen2015highBP,
  title={High-quality sandwiched black phosphorus heterostructure and its quantum oscillations},
  author={Chen, Xiaolong and Wu, Yingying and Wu, Zefei and Han, Yu and Xu, Shuigang and Wang, Lin and Ye, Weiguang and Han, Tianyi and He, Yuheng and Cai, Yuan and others},
  journal={Nature communications},
  volume={6},
  number={1},
  pages={7315},
  year={2015},
  publisher={Nature Publishing Group UK London},
  doi={10.1038/ncomms8315}
}

@article{cao2015quality,
  title={Quality heterostructures from two-dimensional crystals unstable in air by their assembly in inert atmosphere},
  author={Cao, Yang and Mishchenko, A and Yu, GL and Khestanova, Ekaterina and Rooney, AP and Prestat, Eric and Kretinin, AV and Blake, P and Shalom, M Ben and Woods, Colin and others},
  journal={Nano letters},
  volume={15},
  number={8},
  pages={4914--4921},
  year={2015},
  publisher={ACS Publications},
  doi={10.1021/acs.nanolett.5b00648}
}

@article{strukov2008missing,
  title={The missing memristor found},
  author={Strukov, Dmitri B and Snider, Gregory S and Stewart, Duncan R and Williams, R Stanley},
  journal={Nature},
  volume={453},
  number={7191},
  pages={80--83},
  year={2008},
  publisher={Nature Publishing Group UK London},
  doi={10.1038/nature06932}
}

@article{lanza2022memristive,
  title={Memristive technologies for data storage, computation, encryption, and radio-frequency communication},
  author={Lanza, Mario and Sebastian, Abu and Lu, Wei D and Le Gallo, Manuel and Chang, Meng-Fan and Akinwande, Deji and Puglisi, Francesco M and Alshareef, Husam N and Liu, Ming and Roldan, Juan B},
  journal={Science},
  volume={376},
  number={6597},
  pages={eabj9979},
  year={2022},
  publisher={American Association for the Advancement of Science},
  doi={10.1126/science.abj9979}
}

@article{milloch2024mott,
  title={Mott materials: unsuccessful metals with a bright future},
  author={Milloch, Alessandra and Fabrizio, Michele and Giannetti, Claudio},
  journal={npj Spintronics},
  volume={2},
  number={1},
  pages={49},
  year={2024},
  publisher={Nature Publishing Group UK London},
  doi={10.1038/s44306-024-00047-y}
}

@book{shaw1992physics,
  title     = {The Physics of Instabilities in Solid State Electron Devices},
  editor    = {Shaw, Melvin P. and Mitin, Vladimir V. and Sch{\"o}ll, Eckehard and Grubin, Harold L.},
  publisher = {Springer},
  address   = {New York, NY},
  year      = {1992},
  isbn      = {978-1-4899-2346-2},
  doi       = {10.1007/978-1-4899-2344-8}
}

@article{abraham2020astability,
  title={Astability versus bistability in van der Waals tunnel diode for voltage controlled oscillator and memory applications},
  author={Abraham, Nithin and Murali, Krishna and Watanabe, Kenji and Taniguchi, Takashi and Majumdar, Kausik},
  journal={Acs Nano},
  volume={14},
  number={11},
  pages={15678--15687},
  year={2020},
  publisher={ACS Publications},
  doi={10.1021/acsnano.0c06630}
}

@article{shim2016phosphorene,
  title={Phosphorene/rhenium disulfide heterojunction-based negative differential resistance device for multi-valued logic},
  author={Shim, Jaewoo and Oh, Seyong and Kang, Dong-Ho and Jo, Seo-Hyeon and Ali, Muhammad Hasnain and Choi, Woo-Young and Heo, Keun and Jeon, Jaeho and Lee, Sungjoo and Kim, Minwoo and others},
  journal={Nature communications},
  volume={7},
  number={1},
  pages={13413},
  year={2016},
  publisher={Nature Publishing Group UK London},
  doi={10.1038/ncomms13413}
}

@article{kumar2022dynamical,
  title={Dynamical memristors for higher-complexity neuromorphic computing},
  author={Kumar, Suhas and Wang, Xinxin and Strachan, John Paul and Yang, Yuchao and Lu, Wei D},
  journal={Nature Reviews Materials},
  volume={7},
  number={7},
  pages={575--591},
  year={2022},
  publisher={Nature Publishing Group UK London},
  doi={10.1038/s41578-022-00434-z}
}

@article{pazos2025synaptic,
  title={Synaptic and neural behaviours in a standard silicon transistor},
  author={Pazos, Sebastian and Zhu, Kaichen and Villena, Marco A and Alharbi, Osamah and Zheng, Wenwen and Shen, Yaqing and Yuan, Yue and Ping, Yue and Lanza, Mario},
  journal={Nature},
  volume={640},
  number={8057},
  pages={69--76},
  year={2025},
  publisher={Nature Publishing Group UK London},
  doi={10.1038/s41586-025-08742-4}
}

@article{boudou1987hysteresis,
  title={Hysteresis IV effects in short-channel silicon MOSFET's},
  author={Boudou, Alain and Doyle, Brian S},
  journal={IEEE electron device letters},
  volume={8},
  number={7},
  pages={300--302},
  year={1987},
  publisher={IEEE},
  doi={10.1109/EDL.1987.26638}
}

@article{moselund2008punch,
  title={Punch-through impact ionization MOSFET (PIMOS): From device principle to applications},
  author={Moselund, KE and Bouvet, D and Pott, V and Meinen, C and Kayal, M and Ionescu, AM},
  journal={Solid-state electronics},
  volume={52},
  number={9},
  pages={1336--1344},
  year={2008},
  publisher={Elsevier},
  doi={10.1016/j.sse.2008.04.021}
}

@article{dutta2017leaky,
  title={Leaky integrate and fire neuron by charge-discharge dynamics in floating-body MOSFET},
  author={Dutta, Sangya and Kumar, Vinay and Shukla, Aditya and Mohapatra, Nihar R and Ganguly, Udayan},
  journal={Scientific reports},
  volume={7},
  number={1},
  pages={8257},
  year={2017},
  publisher={Nature Publishing Group UK London},
  doi={10.1038/s41598-017-07418-y}
}

@article{song2023recent,
  title={Recent advances and future prospects for memristive materials, devices, and systems},
  author={Song, Min-Kyu and Kang, Ji-Hoon and Zhang, Xinyuan and Ji, Wonjae and Ascoli, Alon and Messaris, Ioannis and Demirkol, Ahmet Samil and Dong, Bowei and Aggarwal, Samarth and Wan, Weier and others},
  journal={ACS nano},
  volume={17},
  number={13},
  pages={11994--12039},
  year={2023},
  publisher={ACS Publications},
  doi={10.1021/acsnano.3c03505}
}

@article{tanoue1988triple,
  title={A triple-well resonant tunneling diode for multiple-valued logic application},
  author={Tanoue, T and Mizuta, Hiroshi and Takahashi, S},
  journal={IEEE electron device letters},
  volume={9},
  number={8},
  pages={365--367},
  year={1988},
  publisher={IEEE},
  doi={10.1109/55.745}
}

@article{tian2026reconfigurable,
  title={Reconfigurable and multifunctional circuits using the Stark effect in black phosphorus},
  author={Tian, He and Hou, Zhan and Wu, Fan and Jiang, Jing-Wen and Wu, Dai-Xuan and Shen, Yang and Xu, Ting-Yi and Xue, Xiao-Yong and Wang, Zi-Ming and Guo, Hao and others},
  journal={Nature Physics},
  pages={1--9},
  year={2026},
  publisher={Nature Publishing Group UK London},
  doi={10.1038/s41567-026-03293-5}
}

@article{wu2012three,
  title={Three-terminal graphene negative differential resistance devices},
  author={Wu, Yanqing and Farmer, Damon B and Zhu, Wenjuan and Han, Shu-Jen and Dimitrakopoulos, Christos D and Bol, Ageeth A and Avouris, Phaedon and Lin, Yu-Ming},
  journal={ACS nano},
  volume={6},
  number={3},
  pages={2610--2616},
  year={2012},
  publisher={ACS Publications},
  doi={10.1021/nn205106z}
}

@article{liu2017gate,
  title={Gate-tunable giant stark effect in few-layer black phosphorus},
  author={Liu, Yanpeng and Qiu, Zhizhan and Carvalho, Alexandra and Bao, Yang and Xu, Hai and Tan, Sherman JR and Liu, Wei and Castro Neto, AH and Loh, Kian Ping and Lu, Jiong},
  journal={Nano letters},
  volume={17},
  number={3},
  pages={1970--1977},
  year={2017},
  publisher={ACS Publications},
  doi={10.1021/acs.nanolett.6b05381}
}

\newpage
\setcounter{figure}{0}
\renewcommand{\thesection}{}
\renewcommand{\thesubsection}{S\arabic{subsection}}
\renewcommand{\theequation} {S\arabic{equation}}
\renewcommand{\thefigure} {S\arabic{figure}}
\renewcommand{\thetable} {S\arabic{table}}

\end{document}